\documentclass[11pt]{jhep3}
\usepackage{amssymb}
\usepackage{bm}
\usepackage{latexsym}
\usepackage{amssymb}
\usepackage{amsmath}
\usepackage{amsfonts}
\newcommand{\UnderscoreCommands}{\do\href}
\usepackage[strings]{underscore}

\renewcommand{\leq}{\leqslant}

\newcommand{\D}[1]{\mathrm{D}{#1}}
\newcommand{\Dbar}[1]{\overline{\D{#1}}}
\newcommand{\Mp}{M_{\mathrm{P}}}
\newcommand{\A}{\mathcal{A}}
\newcommand{\vect}[1]{\bm{\mathrm{{#1}}}}
\newcommand{\e}[1]{\mathrm{e}^{{#1}}}
\newcommand{\im}{\mathrm{i}}
\newcommand{\meff}{m_{\mathrm{eff}}}
\newcommand{\etal}{\emph{et al.}}
\newcommand{\fnl}{f_{\mathrm{NL}}}
\newcommand{\mstar}{m_{\star}}
\newcommand{\gstar}{g_{\star}}
\newcommand{\hstar}{h_{\star}}

\DeclareMathOperator{\Or}{O}

\renewcommand{\d}{\partial}
\newcommand{\flatd}{\mathrm{d}}

	\title{Preheating in Dirac-Born-Infeld inflation}
	\author{Nazim Bouatta, Anne-Christine Davis and Raquel H. Ribeiro\\
		e-mail: \email{N.Bouatta, A.C.Davis and R.Ribeiro@damtp.cam.ac.uk} \\
		Department of Applied Mathematics and Theoretical Physics\\
		Centre for Mathematical Sciences, Wilberforce Road \\
		Cambridge CB3 0WA, United Kingdom} 

	\author{David Seery\\
		e-mail: \email{D.Seery@sussex.ac.uk}\\
			Astronomy Centre, University of Sussex \\
			Falmer, Brighton BN1 9QH, United Kingdom} 

	\abstract{
	We study how the universe reheats following an era of chaotic
	Dirac--Born--Infeld inflation, and compare the
	rate of particle production
	with that in models based on canonical kinetic terms.
	Particle production occurs through non-perturbative
	resonances whose structure is modified by the nonlinearities
	of the Dirac--Born--Infeld action.
	We investigate these modifications and show that
	the reheating process may be efficient.
	We estimate the initial
	temperature of the subsequent hot, radiation-dominated phase.}
	
	\preprint{arXiv:1005.2425}

\begin{document}

	\section{Introduction}
	\label{sec:introduction}
	
	Measurements of the microwave background temperature
	made by the WMAP satellite over the last decade
	\cite{Komatsu:2010fb}
	are consistent with the idea that structure in the universe
	originated as small fluctuations during a primordial era of accelerated
	expansion, or `inflation'. Inflation leaves
	the universe in a cold vacuum state, devoid of radiation and matter.
	If our own universe experienced an inflationary era in its past,
	we must conclude that some process
	repopulated its
	sterile vacuum with the abundant quanta we see today.
	Much less is known about this `reheating' process
	than is understood
	about the observationally accessible density fluctuation.

	Properties of inflationary fluctuations largely decouple from
	the details of whatever physics drives the accelerating epoch,
	leaving only weak traces.%
		\footnote{These traces are presently the subject of
		intense theoretical and observational effort; see, e.g.
		Ref.~\cite{Chen:2010xk}.}
	This freedom has encouraged many proposals
	for the underlying microphysics,
	nearly all of which are equally unconstrained by observation.
	In an especially interesting proposal, due to
	Silverstein \& Tong \cite{Silverstein:2003hf} and later
	elaborated by Alishahiha, Silverstein \& Tong
	\cite{Alishahiha:2004eh},
	inflation occurs while a $\D{3}$-brane
	moves in a warped background spacetime.%
		\footnote{D-branes are extended objects in spacetime.
		For a review of their properties see Johnson
		\cite{Johnson:2003gi};
		inflation in such models has been reviewed by
		McAllister \& Silverstein
		\cite{McAllister:2007bg}. Many detailed properties
		were discussed by
		Chen~\cite{Chen:2005ad}.}
	As we shall discuss in \S\ref{sec:review}, the motion of the brane
	is controlled by an action of Dirac--Born--Infeld (``DBI'') type
	and the resulting model is known as DBI inflation.
	In this paper we study the reheating era which must follow such
	an inflationary phase.
	Reheating in related brane-world models has been studied
	in Refs.~\cite{Cline:2002it, Brodie:2003qv, Dvali:2003ar,
	Frey:2005jk, Chialva:2005zy}.
	In contrast to the primordial perturbations, reheating depends
	strongly on many model-specific details~\cite{Kofman:1996mv,Kaiser:1995fb}.
	It is a subtle function of
	nonlinear physics, probing information complementary to the linear
	physics of the density fluctuation. Nonlinear properties vary widely
	between competing inflationary models,
	and there are
	important differences between
	DBI models and those with canonical kinetic terms.

	Many authors have contributed to the theory of particle production,
	which is now well-developed
	\cite{Traschen:1990sw,Kofman:1994rk,Shtanov:1994ce,Kofman:1997yn}.%
		\footnote{The theory of preheating has been reviewed
		by Allahverdi {\etal} \cite{Allahverdi:2010xz},
		Kofman {\etal} \cite{Kofman:2008zz}
		and Bassett {\etal} \cite{Bassett:2005xm}.
		See also Ref.~\cite{Kohri:2009ac}.}
	In \emph{perturbative reheating}, the decay rate of inflaton particles
	into other species of matter
	is calculated directly from an S-matrix.
	These calculations assume any relics of previous decays to be
	sufficiently diffuse that particle production takes place into
	an effectively pristine vacuum.
	It was later understood that a Bose enhancement of the decay rate
	may occur for integer-spin species
	if the products of previous decays accumulate, leading
	to a resonant phase of out-of-equilibrium production known as
	\emph{preheating}.
	A different effect, so-called \emph{tachyonic
	preheating} \cite{Felder:2000hj,Felder:2001kt, Dufaux:2006ee}, is
	associated with the end of inflation in
	certain models.
	Unlike conventional preheating this does not merely
	convert one species
	of particle into another, but (like any tachyonic instability)
	reflects a preference to
	convert one vacuum state to another
	by the rapid accumulation and condensation of particles.

	These phenomena may occur in any effective field theory
	of the post-inflationary universe, whatever its microphysical origin.
	The general theory has subsequently been applied to concrete models,
	including some versions of brane inflation,
	in which it is possible to give a more refined interpretation
	\cite{Kachru:2003sx,Barnaby:2004gg}.
	In these models, inflation ends when the moving $\D{3}$-brane
	becomes close to an antibrane (a $\Dbar{3}$-brane), allowing
	open strings stretching between the
	$\D{3}$--$\Dbar{3}$ pair
	to become excited.
	The open string states include a tachyon which induces
	fragmentation of the $\D{3}$ brane into $\D{0}$ branes,
	described on the $\D{3}$ worldvolume by
	a phase of tachyonic preheating.
	The $\D{0}$ fragments decay into closed string states,
	which are enumerated by the
	Kaluza--Klein (``KK'') modes of supergravity fields in the warped
	background. These KK modes subsequently decay
	into the species of
	matter and radiation which populate our universe
	\cite{Barnaby:2004gg}.
	
	In this paper we study a version of the DBI model in which
	the end of inflation is less dramatic.
	We work in a model where
	the $\D{3}$ comes to rest near an extremity of its warped
	background, referred to as the ``tip of the throat.''
	As it settles on its resting-place it executes coherent
	oscillations, inducing
	growing fluctuations which fold and wrinkle its surface.
	This scenario was suggested by Silverstein \& Tong
	\cite{Silverstein:2003hf}.
	Our interest is in the \emph{qualitative} differences which
	may emerge in concrete models. Should we expect the reheating
	temperature to increase or decrease? Is parametric resonance
	more efficient, perhaps due to the extra non-linearities of the
	DBI action, or do these same non-linearities conspire to shut off
	the resonant channels?
	For this reason we focus on differences with the case
	of  canonical
	kinetic terms and largely ignore similarities. For example,
	the use of perturbation theory in either case will break down at the onset
	of back-reaction, when the population of matter species grows to the
	point where it cannot be ignored.
	This affects the canonical and DBI cases equally, and we restrict our
	comparison to times before back-reaction becomes significant.

	This paper is organized as follows.
	In \S\ref{sec:review} we review the dynamics of inflation in DBI models and
	discuss a toy model in which inflation ends as the brane
	settles to its final position in the warped background.
	This model is comparable to chaotic inflation in a theory with
	canonical kinetic terms.
	In \S\ref{sec:res} we discuss parametric resonance,
	making a comparison with canonical chaotic inflation.
	In \S\ref{sec:pr} we study the stage of reheating and
	estimate the reheating temperature.
	We conclude in \S\ref{sec:conclude}.
	
	Throughout this paper, we work in units where $\hbar = c = 1$
	and set the reduced Planck mass to unity,
	$\Mp \equiv (8 \pi G)^{-1/2} = 1$.
	The metric signature is taken to be $(-, +, +, +)$.

	\section{Dirac--Born--Infeld inflation}
	\label{sec:review}
	
	In \S\ref{sec:dbi} we
	review inflation using a Dirac--Born--Infeld Lagrangian
	and study the oscillating era which occurs while the brane
	is settling
	near the tip of throat.
	In \S\ref{sec:coupling}
	we discuss possible couplings of the inflaton to matter.

	\subsection{Review of DBI Cosmology}
	\label{sec:dbi}

	In the scenario discussed by Silverstein \& Tong
	\cite{Silverstein:2003hf}, one or more $\D{3}$-branes
	move in a warped throat, taken to be an elongated
	region of a six-dimensional compact manifold
	with metric $\flatd s_6^2$. Transverse distance within the
	throat is measured by a parameter $\tau$,
	and the entire ten-dimensional
	geometry can be written schematically
	\begin{equation}
		\flatd s_{10}^2 = h^{-1/2}(\tau) \eta_{\mu\nu}
		\flatd x^\mu \flatd x^\nu +
		h^{1/2}(\tau) \flatd s_6^2 ,
		\label{eq:metric}
	\end{equation}
	where $h(\tau)$ is a warp factor.
	As $\tau \rightarrow \infty$ there is an ultraviolet regime
	where the throat is glued onto to the rest of the $\flatd s_6^2$.
	In this region
	the throat corresponds to a cone over an Einstein manifold.
	In the infrared regime where
	$\tau \rightarrow 0$,
	corresponding to the ``tip'' of the throat,
	fluxes wrapped along cycles of the conifold smooth
	the conical singularity with an $S^3$ cap.
	
	A low energy description of a $\D{3}$ brane travelling in this
	warped throat is given by
	\begin{equation}
		S = \frac{1}{2} \int \flatd^4 x \; \sqrt{-g} \;
		\Big\{
			R + 2 P(\phi, X)
		\Big\} ,
		\label{eq:action}
	\end{equation}
	where $g_{\mu\nu}$ is the pull-back
	of Eq.~\eqref{eq:metric} to the brane worldvolume,
	and $R$ is the Ricci scalar constructed from $g_{\mu\nu}$.
	The worldvolume scalar $\phi$ is a collective coordinate which
	measures the radial position of the brane within the throat,
	and
	$X \equiv - g^{\mu \nu} \d_\mu \phi
	\d_\nu \phi$.
	The action describing evolution of $\phi$ satisfies
	\begin{equation}
		P(\phi, X) = - f_1(\phi) \left[ \sqrt{1 - f_2(\phi) X} - 1 \right]
		- V(\phi) ,
		\label{eq:DBI}
	\end{equation}
	in which $f_1$ and $f_2$ are determined by the warped geometry
	and $V$ is a potential
	which describes
	how the brane is attracted towards the bottom of the throat.
	By choosing coordinates we can arrange that $f_1 = f_2^{-1} \equiv f^{-1}$.
	In these coordinates $f^{-1}$ measures the tension of the
	$\D{3}$-brane, redshifted by the warp factor,
	giving $f$ canonical dimension $[\mathrm{GeV}^{-4}]$.
	The proper tension $T_3$ is set in terms of
	the inverse string tension $2\pi \alpha'$
	and the string coupling,
	$g_s$, by $T_3^{-1} = (2\pi)^3 (\alpha')^2 g_s$.
	After redshifting, $f = h/T_3 $.

	If the throat is exactly
	anti de Sitter, the result is a static BPS state
	in which the Coulomb
	potential is quartic.
	Conformal invariance forbids the existence of a mass term.
	If the throat is cut off then conformal invariance is broken,
	and we expect a mass term to be generated
	which receives contributions from moduli stabilization and bulk
	fluxes
	\cite{Kachru:2003sx,Baumann:2006th,Baumann:2007np,Baumann:2007ah,
	Krause:2007jk,
	Baumann:2009qx,Baumann:2010sx}.
	In a general compactification, a spectrum of terms will be generated
	including a $\phi^{3/2}$ term which may lead to
	a phase of `ledge' or inflexion-point
	inflation.
	We assume this term is absent, or forbidden by a symmetry.
	Therefore, for small departures from the equilibrium point,
	the potential attracting the brane towards the infrared
	tip of the throat is quadratic \cite{Silverstein:2003hf},
	\begin{equation}
		V(\phi) = \frac{1}{2} m^2\phi^2 .
		\label{eq:pot}
	\end{equation}
	D-brane inflation with this potential is sometimes referred to
	as the `UV model'.
	A related
	small-field (`IR') model due to Chen \cite{Chen:2005ad}
	takes $V = V_0 - m^2 \phi^2 / 2$
	with $V_0$ dominant, but in this
	paper we do not consider this possibility.
	In DBI models there is less
	need to tune the contributions to $m^2$ to avoid an
	$\eta$ problem \cite{Easson:2009kk},
	since inflation can be supported on steeper potentials
	than would be allowed with canonical kinetic terms.

	As the brane drifts
	towards the infrared tip of the throat,
	$\phi$ slowly decreases to zero.
	In this limit $\dot{\phi}^2 \ll 1$
	and the square root can be expanded in powers of $X$,
	reproducing the action for
	a canonically normalized scalar field with potential $V$.
	On the other hand,
	if the brane is moving relativistically then $X \sim 1$
	and the square root in Eq.~\eqref{eq:DBI} cannot be reduced.%
		\footnote{In this limit we retain an infinite set of
		non-renormalizable terms involving powers of
		$(\partial \phi)^2$ with coefficients in
		pre-determined ratios, but discard all operators of
		the form $\partial^n \phi$ for $n > 1$. This is nevertheless
		sensible because $\sqrt{1 - f X}$ is the unique operator
		containing only $\partial \phi$ which is
		compatible with higher-dimensional isometries of the
		DBI model, and therefore satisfies a form of the non-renormalization
		theorem \cite{Silverstein:2003hf,Leblond:2008gg,Shandera:2008ai,
		deRham:2010eu}.}
	To measure the relative importance of these nonlinear terms
	it is conventional to define a Lorentz factor
	for the brane,
	\begin{equation}
		\gamma^{-1} \equiv \sqrt{1 - f(\phi) \dot{\phi}^2} .
		\label{eq:lorentz}
	\end{equation}
	Unitarity of Eq.~\eqref{eq:action}
	requires $\gamma$ to be real, and therefore
	$0 \leq f \dot{\phi}^2 < 1$
	\cite{Kabat:1999yq,Silverstein:2003hf},
	which gives a limiting velocity for $\phi$.
	For a slowly drifting brane we have $\gamma \sim 1$, whereas
	relativistic motion corresponds to $\gamma \gg 1$.
	The equations of motion for the inflaton are
	\begin{equation}
		\ddot\phi
		+ 3H \frac{\dot\phi}{\gamma^2}
		- \frac{f'}{f^2}
		+ \frac{3}{2}\frac{f'}{f}\dot{\phi}^2
		+ \frac{1}{\gamma^3} \left(
			\frac{f'}{f^2}
			+ V'
		\right)
		= 0 ,
		\label{eq:inflaton}
	\end{equation}
	where a prime $'$ denotes a derivative with respect to $\phi$.
	In the non-relativistic limit $f \rightarrow 0$ and
	$\gamma \rightarrow 1$, and Eq.~\eqref{eq:inflaton}
	reduces to the Klein--Gordon equation for a canonically normalized
	scalar field.

	Maps of the microwave background temperature anisotropy place
	stringent constraints on $\gamma$, which can source a large
	signal in the equilateral mode of the bispectrum
	of order $\fnl \sim - 0.32 \gamma^2$ \cite{Alishahiha:2004eh}.
	An analysis of WMAP 5-year data
	yields $\fnl = 155 \pm 140$ in this channel
	\cite{Senatore:2009gt}, where the quoted error is $1\sigma$;
	at $95\%$ confidence the requirement is roughly $\gamma \lesssim 18$.
	In this paper we will assume this upper limit will continue to be eroded,
	perhaps ultimately constraining $\gamma$ to be rather small,
	so that it is natural to imagine the model is close to its
	non-relativitic limit.
	Accordingly
	we suppose that inflation is approximately of the slow-roll variety.
	In this limit
	the evolution of $\phi$ is close to a canonical model,
	up to small corrections controlled by the
	deviation of $\gamma$ from unity.
	We will solve Eqs.~\eqref{eq:inflaton}--\eqref{eq:pot}
	perturbatively in powers of these small deviations.
	
	Inflation ends near the tip of the throat,
	where the $S^3$ cap smoothly cuts off
	further decrease of the warp
	factor \cite{Klebanov:2000hb,Kecskemeti:2006cg}.
	In this region there is no redshifting associated with
	motion down the throat, and the brane tension $f^{-1}$ becomes
	approximately constant.
	Silverstein \& Tong \cite{Silverstein:2003hf} suggested that
	the relevant physics of the capped throat could be captured by
	approximating the tension as
	\begin{equation}
		f(\phi) \equiv \dfrac{\lambda}{\left(\phi^2+\mu^2\right)^2} ,
		\label{eq:tension}
	\end{equation}
	where $\lambda$ is the 't Hooft coupling and
	$\mu$ is the energy scale of the
	infrared cut-off.
	We have chosen the origin of $\phi$ so that
	the final resting place of the brane can be taken to be $\phi = 0$.
	The warp factor corresponding to Eq.~\eqref{eq:tension}
	is not itself a solution of Einstein's equations, but
	is qualitatively reasonable.
	When $\phi \ll \mu$, this becomes independent of the brane position
	$\phi$,
	\begin{equation}
		f \equiv \frac{\lambda}{\mu^4} ,
		\label{eq:cutoff}
	\end{equation}
	We will discuss natural values
	for $\lambda$ and $\mu$ in \S\ref{sec:pr}.
	
	We restrict our attention to small oscillations around this
	position, contained entirely within the cut-off region where $f^{-1}$
	is approximately constant, and suppose that $m \gg H$.
	Therefore many oscillations take place in one Hubble time
	and terms in Eq.~\eqref{eq:inflaton} arising from Hubble friction
	can be ignored.
	We will justify these statements below.
	Making all these approximations and taking
	the non-relativistic limit where $\gamma = 1$, coherent
	oscillations of the brane
	occur with constant amplitude $\A$,
	\begin{equation}
		\phi_0(t) \simeq \A \cos mt .
		\label{solhomogene}
	\end{equation}

	\paragraph{Validity of perturbative approximation.}
	Our analysis depends on the approximations described above, and will
	provide a useful description only when perturbative relativistic
	corrections dominate those from other sources.

	When is this likely to occur?
	After integrating out fluxes and other UV perturbations,
	we are left with a low energy theory controlled by
	an effective potential for the D-brane \cite{Baumann:2009qx,Baumann:2010sx},
	of the form $V = \sum_i c_i m^{4-\Delta_i} \phi^{\Delta_i}$,
	where the $c_i$ are expected to be $\Or(1)$ in a generic compactification.
	The mass $m$ is an ultraviolet scale which determines where the
	throat is glued onto the main bulk of the compactification manifold.
	The leading term corresponds to $\Delta = 3/2$.
	Assuming some symmetry forces this to be absent,
	the next $\Delta = 2$
	term will typically dominate higher contributions
	when $\phi \ll m$.
	
	The infrared cutoff scale, $\mu$, must satisfy $\mu < m$.
	Therefore, provided we take $\phi < \mu$, in order that we are
	safely within a region where the warp factor is constant,
	we can expect corrections from both features in the throat geometry
	and higher-order terms in the D-brane potential to be negligible.
		
	Now consider our neglect of Hubble drag in
	Eq.~\eqref{solhomogene}.
	In practice, Hubble drag would cause $\A$ to decrease slowly,
	roughly like $(mt)^{-1}$ \cite{ArmendarizPicon:2007iv}.
	Suppose $H_f$ and $H_i$ are two values of the Hubble parameter
	separated by $N$ oscillations. Since the brane worldvolume experiences
	a matter-dominated era while the coherent oscillations are
	ongoing, these values are related by
	\begin{equation}
		\frac{H_i}{H_f} \sim 1 + A N \frac{H_i}{m} ,
		\label{eq:hi}
	\end{equation}
	where $A$ is a constant of order unity.
	The corresponding oscillation amplitudes are
	\begin{equation}
		\frac{\A_i}{\A_f} \sim 1 + A' N \frac{\A_i}{\Mp} ,
		\label{eq:Ai}
	\end{equation}
	in which $A'$ is a different constant, also of order unity.
	Unless $N \gg 1$,
	our assumption that $m \gg H$ is sufficient to ensure
	$H_i \approx H_f$.
	In addition,
	Eq.~\eqref{eq:Ai} shows that $\A_i \ll \Mp$ is sufficient to ensure
	$\A_i \approx \A_f$.
	In view of the requirement $\A \ll m$
	required to ensure the
	quadratic potential is an adequate approximation, where $m$ is typically
	of order the infrared cutoff scale,
	this is amply satisfied.
	Therefore, in what follows, we will
	restrict our attention to $N < 10^2$ oscillations
	and neglect the time dependence of $H$ and $\A$.
	
	\paragraph{Relativistic corrections to background solution.}
	Corrections to the non-relativistic motion
	described by~\eqref{solhomogene}
	will be perturbative provided
	$|f \dot{\phi}_0^2 | \ll 1$.
	Introducing a dimensionless positive quantity $\epsilon$, satisfying
	\begin{equation}
		\epsilon \equiv f m^2 \A^2 ,
		\label{eq:defepsilon}
	\end{equation}
	this perturbative condition can be rewritten $\epsilon \ll 1$.
	We must nevertheless work in a regime where $\epsilon \gtrsim
	\phi/m$ or $\mu/m$, in order that the small
	relativistic corrections we obtain
	are not dominated by others which we do not calculate.
	We should trust our analysis for perhaps $\epsilon \sim 10^{-1}$
	to $10^{-3}$, but not for significantly larger or smaller $\epsilon$.

	To go further, we must determine the
	leading $\Or(\epsilon)$ corrections to~\eqref{solhomogene}.
	Note that $\epsilon$ is not the conventional slow-roll
	parameter, given by~$- \dot{H}/H^2$, and these corrections are not
	related to the slow-roll approximation.
	The governing equation can be obtained by dropping Hubble
	friction terms in Eq.~\eqref{eq:inflaton}, yielding
	\begin{equation}
		\ddot{\phi}
		+ \left(
			1 - \frac{3}{2}f\dot{\phi}^2
		\right) \frac{\d V}{\d\phi}
		\simeq 0 ,
		\label{approxeq}
	\end{equation}
	It is tempting to imagine that Eq.~\eqref{approxeq} can be solved
	perturbatively, 
	writing $\phi = \phi_0 + \delta \phi$ with
	$\delta \phi$ of order $\Or(\epsilon)$.
	However, Eq.~\eqref{approxeq} is rather nonlinear
	and also induces corrections to the fundamental frequency
	$m$.
	To capture these effects, consider a WKB-type solution
	of the form $\phi = \A \cos \theta(t)$
	and solve for the phase $\theta(t)$
	to order $\Or(\epsilon)$.
	To match smoothly with~\eqref{solhomogene} in the limit
	$\epsilon \rightarrow 0$,
	we must have $\theta(t) = m t + \delta \theta(t)$.
	From Eq.~\eqref{approxeq} we conclude that $\delta \theta(t)$
	must satisfy
	\begin{equation}
		\delta \ddot{\theta} +
		2 m (\cot mt)  \delta \dot{\theta} +
		\frac{3 m^2 \epsilon}{2} \sin m t \,  \cos m t = 0 .
	\end{equation}
	Dropping an unwanted divergent contribution,%
		\footnote{This mode is proportional to
		$\cot mt$, which is divergent at $t=0$.
		It is projected out by the boundary conditions,
		and therefore does not lead to an instability.}
	the solution is
	\begin{equation}
		\phi(t) = \A \cos \left\{
			m t \left( 1 - \frac{3 \epsilon}{16} \right) +
			\frac{3 \epsilon}{32} \sin 2 m t
		\right\} .
		\label{eq:fullphi}
	\end{equation}
	Corrections to the amplitude are also calculable in principle,
	but need not be computed if we are ignoring the time dependence
	of $\A$.

	In comparison with the canonical solution, the phase
	shows a secular drift proportional to $\epsilon m t$.
	Although the calculation remains under perturbative control,
	this drift generates order unity
	deviations from the non-relativistic background
	after $1/\epsilon$ oscillations.
	There is also a superposed phase oscillation,
	proportional to $|\epsilon \sin 2 m t| \ll 1$,
	which always generates small corrections.
	To make contact with the canonical solution,~\eqref{eq:fullphi}
	can be written in the alternate form
	\begin{equation}
		\phi(t) =
		\A
		\left\{
			1 - \frac{3 \epsilon}{16} \sin^2 \left[m t
			\left( 1 - \frac{3 \epsilon}{16} \right)\right]
		\right\}
		\cos m t \left( 1 - \frac{3 \epsilon}{16} \right) ,
		\label{eq:phi}
	\end{equation}
	which is equivalent to~\eqref{eq:fullphi}
	up to $\Or(\epsilon)$.
	Eq.~\eqref{eq:phi}
	exhibits the first perturbative relativistic corrections
	to the motion of the brane:
	first, a frequency shift
	in the background motion $\phi_0$, corresponding to
	$m \mapsto m\,(1 - 3 \epsilon/16)$;
	and second, a small periodic fluctuation superposed
	over the background oscillations.

	\subsection{Coupling to matter fields}
	\label{sec:coupling}

	The reheating phase is designed to
	fill the universe with Standard Model degrees of freedom.
	Pauli's exclusion principle usually prevents efficient preheating
	into fermions \cite{Greene:2000ew}, so we suppose the
	inflaton $\phi$ couples to bosonic modes
	which are taken to have spin zero.
	The resulting scenario is frequently adopted in the literature, and
	gives an acceptable phenomenology
	subsequently decay to fermions is efficient.
	We treat the couplings between the inflaton and the matter
	modes phenomenologically, denoting the matter field
	$\chi$ and taking it to be described by the low energy
	effective Lagrangian
	\begin{equation}
		\mathcal{L} \supseteq 
		- \frac{1}{2} \d_\mu \chi\d^\mu \chi
		- \frac{1}{2} m_\chi^2 \chi^2
		- \frac{1}{2} g^2 \phi^2 \chi^2
		\label{masterequation} ,
	\end{equation}
	where $g$ is a dimensionless coupling
	and $m_\chi$ is a bare mass.
	The symbol `$\supseteq$' is used to indicate that the Lagrangian
	contains these contributions among others.
	We do not claim that such interactions necessarily
	arise in string models, although we hope the resulting phenomenology
	will be representative of those that do.
	The equation of motion for $\chi$ is
	\begin{equation}
		\ddot{\chi} + 3 H \dot{\chi} - \frac{1}{a^2} \nabla^2 \chi
		+ \left(
			m_\chi^2 + g^2 \phi^2
		\right) \chi = 0 ,
		\label{mattereq1}
	\end{equation}
	where $\nabla^2$ is the spatial coordinate Laplacian.
	Matter production in this model is studied in \S\ref{sec:matter}.
	
	In \S\ref{sec:sym-break}
	we will consider an alternative model,
	in which the matter field $\chi$ is coupled to
	a symmetry-breaking potential of the form
	\begin{equation}
		V(\phi, \chi) = \dfrac{1}{2}m^2\left(\phi-\sigma\right)^2 
		+
		\dfrac{1}{2} g^2\phi^2\chi^2 .
		\label{eq:pot-sym}
	\end{equation}
	This causes the brane to be
	attracted to a radial position for which $\phi = \sigma$,
	representing a toy model of compactifications in which
	the mobile D-branes are brought to rest at a finite point in the
	throat.

	\section{Parametric resonance and particle production}
	\label{sec:res}

	As the brane settles towards its final resting place,
	its coherent oscillations
	lead to copious non-perturbative particle production.
	We briefly review this process, with the aim of fixing our notation.
	
	To quantize any field, such as
	the inflaton $\phi$ or the matter field $\chi$,
	we pick two independent
	solutions to the mode equation,
	\begin{equation}
		\ddot{\chi}_k + 3 H \dot{\chi}_k +
		\left(
			k^2 +m_\chi^2+ g^2 \phi^2
		\right) \chi_k = 0
		\label{fouriermode} ,
	\end{equation}
	where $k$ is the comoving wavenumber,
	and to be concrete we have chosen to frame our discussion for
	$\chi$ although an equivalent construction can be given for $\phi$.
	Label these solutions $\chi_k^{\mathrm{cl}}$ and
	$\chi_k^{\mathrm{cl}\ast}$.
	The Heisenberg field corresponding to $\chi$ can be
	written
	\begin{equation}
		\chi(t, \vect{x})
		=
		\int \frac{\flatd^3 k}{(2\pi)^3}
		\left(
			\chi_k^{\mathrm{cl}} a^\dag_k
			+ \chi_k^{\mathrm{cl}\ast} a_{-k}
		\right)
		\e{\im \vect{k} \cdot \vect{x}} ,
	\end{equation}
	in which we have assumed that the normalisation of
	$\{ \chi_k^{\mathrm{cl}}, \chi_k^{\mathrm{cl}\ast} \}$
	has been adjusted so that
	$a_k$ and $a^\dag_k$ obey the usual creation--annihilation algebra,
	\begin{equation}
		[ a_k , a^\dag_{k'} ] = (2\pi)^3 \delta( \vect{k} - \vect{k}' ) .
	\end{equation}
	What is the occupation number $n_k$ for momenta of magnitude $k$?
	The energy density, $E_k$,
	associated with $\chi$-quanta of wavenumber $k$
	satisfies $2 E_k = | \dot{\chi}_k^{\mathrm{cl}} |^2
	+ \omega_k^2 | \chi_k^{\mathrm{cl}} |^2$,
	where $\omega_k^2 = k^2 + g^2 \phi^2$.
	Accounting for subtraction of the energy density associated with
	vacuum fluctuations, when the occupation number is large it must be
	given to a good approximation by
	\begin{equation}
		n_k \approx \frac{E_k}{\omega_k}
		= \frac{\omega_k}{2} \left(
			\frac{| \dot{\chi}_k^{\mathrm{cl}} |^2}{\omega_k^2} +
			| \chi_k^{\mathrm{cl}} |^2
		\right) - \frac{1}{2} .
		\label{eq:occupy}
	\end{equation}
	It follows that the occupation number
	can be determined from knowledge of
	the mode functions $\chi_k^{\mathrm{cl}}$ and their derivatives;
	when $|\chi_k^{\mathrm{cl}}|$ or
	$|\dot{\chi}_k^{\mathrm{cl}}|$ are large, we expect
	large occupation numbers.
	In what follows we shall not need to make use of the quantum field
	$\chi$, which allows us to drop the label `cl' on each mode function.
	In particular, if $\chi_k$ experiences exponential growth---implying
	that the production of $\chi$ particles is stimulated
	by the presence of existing $\chi$-quanta---then
	Eq.~\eqref{eq:occupy} shows we can expect
	$n_k$ to attain macroscopic values very rapidly.
	This is the phenomenon of preheating, discussed
	in \S\ref{sec:introduction}.
	
	To determine when $\chi_k$ grows, it is often useful to
	translate the mode equation for $\chi_k$
	into a version of Mathieu's equation or its generalizations.
	The classical Mathieu equation is
	\cite{McLachlan,Arscott}
	\begin{equation}
		\frac{\flatd^2 \chi_k}{\flatd z^2}
		+ (A_k - 2q \cos 2z) \chi_k = 0 ,
		\label{mathieu}
	\end{equation}
	where we have introduced a dimensionless time $z$
	to be defined below.
	In the following sections
	we give explicit expressions for $A_k$ and $q$
	applicable to both inflaton and matter fluctuations.
	For certain combinations of these parameters
	the solutions of Mathieu's equation exhibit exponential growth.
	If $q \gtrsim A_k$, the effective mass can become temporarily zero.
	In this case it is energetically inexpensive to
	copiously produce $\chi$-quanta, referred to as
	\emph{broad resonance}. Alternatively,
	if $q \ll A_k$ then the effective mass never passes through zero,
	and the combinations which yield exponential growth occur only in
	narrow bands of momentum space.
	This scenario is known as \emph{narrow resonance}.
	These bands are centred on momenta for which
	$A_k = \ell^2$, where $\ell \in \mathbb{N}$,
	and roughly correspond to values of $k$ for which
	$|A_k - \ell^2| \lesssim q^\ell$.
	If $|q| \ll 1$ then at large $\ell$
	these bands become increasingly narrow.

	\subsection{Inflaton production}
	\label{sec:inf-prod}
	
	The foregoing analysis applies equally to inflaton and matter
	fluctuations.
	First, consider the inflaton mode of wavenumber $k$.
	We set $\phi$ equal to its background value,
	denoted $\varphi$,
	which is
	determined up to first-order relativistic corrections by~\eqref{eq:phi}.
	At the onset of oscillations it is reasonable to assume
	that $\chi$ is in its vacuum state $\chi = 0$.
	Until the accumulation of $\chi$-quanta
	causes the background $\chi$ field to grow,
	the inflaton fluctuation 
	$\delta\phi_k$ evolves according to the Klein-Gordon equation,
	to linear order in $\epsilon$,
	\begin{equation}
		\delta \ddot{\phi}_k - 3 \frac{f m^2 \varphi \dot{\varphi}}{\gamma}
		\delta \dot{\phi}_k + \left(
			\frac{m^2}{\gamma^3} + k^2
		\right) \delta \phi_k = 0 .
		\label{eq:infk}
	\end{equation}
	We expand uniformly to leading order in $f \dot{\varphi}^2$
	and make the change of variables $\delta \phi_k = \xi y_k$,
	choosing $\xi$ to eliminate the friction term in~\eqref{eq:infk},
	\begin{equation}
		\frac{\flatd \ln \xi}{\flatd t} = \frac{3}{2} f m^2 \varphi
			\dot{\varphi} .
	\end{equation}
	The solution to this differential equation is
	\begin{equation}
		\xi = \xi_0 \exp \left\{
			\frac{3}{2} f m^2 \left[ \varphi^2(t) - \varphi^2(t_0) \right]
		\right\} ,
	\label{eq:xisolution}
	\end{equation}
	where $\xi = \xi_0$ at $t = t_0$.
	This is an effective friction, or `braking,' which
	could be interpreted as a perturbative
	manifestation of the D-cceleration mechanism
	\cite{Silverstein:2003hf}.
	It is a novel effect in the DBI scenario, with no analogue
	for canonical kinetic terms. Accordingly, the braking disappears
	in the limit $\epsilon \rightarrow 0$.
	The remaining time dependence
	is carried by $y_k$, which satisfies a Mathieu equation with $A_k$
	and $q$ determined, to leading order in $\epsilon$, by
	\begin{equation}
		A_k = \left(1 + \frac{k^2}{m^2} \right)
			\left(1 + \frac{3 \epsilon}{8} \right)
			- \frac{3 \epsilon}{4}
		\quad \mbox{and} \quad
		q = \frac{3 \epsilon}{8} .
	\end{equation}
	The associated dimensionless time $z$,
	which was used to write Eq.~\eqref{mathieu}, satisfies
	$z = m t ( 1 - 3 \epsilon / 16 )$.
	Evidently $q \ll A_k$, giving resonance only in narrow bands.
	The $\ell = 1$ band is centred at momentum $k_1$, given by
	\begin{equation}
		k_1 = \frac{m}{2} \sqrt{\frac{3 \epsilon}{2}} .
	\end{equation}
	According to Floquet's theory
	each mode behaves like $\e{\mu_k z}$.
	In the $\ell = 1$ band,
	the Floquet exponent $\mu_k$ has positive real solutions
	which correspond
	to exponential growth \cite{McLachlan}
	\begin{equation}
		\mu_k^2 \simeq \left(
			\frac{q}{2}
		\right)^2 -
		\left( A_k^{1/2} - 1 \right)^2 .
	\label{eq:mumathieu}
	\end{equation}
	This exponent is maximal for the central value $k = k_1$
	and decreases for larger or smaller $k$,
	forming a band of width $\Delta_1$,
	\begin{equation}
		\Delta_1 = \frac{m \sqrt{3 \epsilon}}{2}
		\frac{\sqrt{2} - 1}{\sqrt{2}} .
	\end{equation}
	At the centre of the band, the Floquet exponent
	takes the value $\mu_1$
	\begin{equation}
		\mu_1 = \max_{\ell = 1 \; \mathrm{band}} \mu_k
		= \frac{3 \epsilon}{16} .
	\end{equation}
	
	Up to this point, we have written our formulae in terms of the mass,
	$m$, and coupling, $g$, which appear in the Lagrangian.
	However, these quantities only parametrize the theory.
	To make a meaningful comparison between the DBI model and the
	case of canonical kinetic terms, we must express our answer in
	terms of a measurable mass and coupling. We denote these
	measurable quantities $\mstar$ and $\gstar$, to distinguish them from
	the Lagrangian parameters. In the DBI theory the Lagrangian parameters
	have no direct physical significance, whereas in the case of canonical
	kinetic terms they coincide with the measurable mass
	and coupling at tree-level.

	We determine the relationships $\mstar(m)$ and
	$\gstar(g)$ in Appendix~\ref{appendix:canonicnorm}.
	Collecting
	the results of Eqs.~\eqref{eq:xisolution} and~\eqref{eq:mumathieu},
	and expressing the answer in terms of $\mstar$ and $\gstar$,
	we find that
	up to an overall normalization
	near the centre of the resonant band, the
	inflaton modes grow like
	\begin{equation}
		\delta \phi_k \sim
		\exp\left\{
			\dfrac{3\epsilon}{8} \cos
			\left[
				2 \mstar t \left(1-\dfrac{7\epsilon}{16} \right)
			\right]
			+ \dfrac{3\epsilon}{16} \mstar t
		\right\} .
		\label{eq:resultinflaton}
	\end{equation}
	The undetermined normalization absorbs the constant
	$\xi_0$ in~\eqref{eq:xisolution}.

	At early times, the DBI friction and the resonant growth compete,
	and the argument of the exponential in Eq.~\eqref{eq:resultinflaton}
	can even become negative.
	After a few oscillations, however, the cosine which represents the
	friction effect becomes negligible
	in comparison with the resonant term.

	\subsection{Matter particle production}
	\label{sec:matter}

	Our ultimate interest lies with the production of matter particles.
	Consider the potential~\eqref{masterequation}.
	For any particle species which gains a mass
	at energies much less than the inflationary scale
	it will typically be a good approximation
	to assume $m_\chi$ is negligible in comparison with $g \phi$
	unless the amplitude of oscillations is extremely small.
	Therefore each
	$\chi$-mode receives an effective mass
	of order $k^2 + g^2 \phi^2$.
	As in \S\ref{sec:inf-prod} we assume that preheating begins
	in the vacuum $\chi = 0$, which will be valid until
	the accumulating $\chi$-quanta backreact on the zero mode.
	In this vacuum, small fluctuations
	grow according to a Hill equation of type
	\begin{equation}
		\delta \chi''_k +
		\left(
			\theta_0 + 2 \theta_2 \cos 2z + 2 \theta_4 \cos 4z
		\right) \delta \chi_k = 0 ,
	\end{equation}
	where the parameters $\theta_0$, $\theta_2$ and $\theta_4$ obey
	\begin{equation}
		{
		\renewcommand{\arraystretch}{2}
		\begin{array}{lll}
			\theta_0 & = & \displaystyle
			\frac{k^2}{m^2} \left( 1 + \frac{3 \epsilon}{8} \right)
			+ \frac{g^2 \A^2}{2m^2} \left( 1 + \frac{9\epsilon}{32} \right) \\
			\theta_2 & = & \displaystyle
			\frac{g^2 \A^2}{4m^2} \left( 1 + \frac{3 \epsilon}{8} \right) \\
			\theta_4 & = & \displaystyle
			\frac{3}{128} \frac{\A^2}{m^2} \epsilon g^2 ,
		\end{array}
		}
	\end{equation}
	and (as in \S\ref{sec:inf-prod})
	a prime $'$ denotes a derivative with respect
	to $z = mt(1 - 3 \epsilon / 16 )$.
	The solutions of Hill's equation grow and decay with characteristic
	Floquet exponent $\mu_k$ in a way entirely analogous to
	the solutions of Mathieu's equation.
	The exponentially growing solutions organize themselves into
	bands, characterized by integers $\ell$.
	Using the arguments of Floquet theory one can
	obtain an estimate of $\mu_k$ for the $\ell = 1$ band
	(see Appendix~\ref{appendix:Hill}),
	\begin{equation}
		\mu_k^2 \approx
		\left( \frac{\theta_2}{2} \right)^2
		- \left( \theta_0^{1/2} - 1 \right)^2
		\mp \frac{\theta_2}{3} \left( \frac{\theta_4}{2} \right)^2
		\mp \frac{\theta_2^3}{2} .
	\end{equation}
	We believe this expression has not previously appeared in the
	preheating literature.
	In the region where our approximations are valid, we have
	$\A \ll m$.
	Therefore, the terms involving $\theta_4^2$ and $\theta_2^3$
	are negligible.
	Under these conditions
	the Floquet exponent can be simplified, yielding
	\begin{equation}
		\mu_k^2 \approx \left( \frac{\theta_2}{2} \right)^2 -
			\left( \theta_0^{1/2} - 1 \right)^2 \ \ .
	\end{equation}
	The centre of the $\ell = 1$ band lies at $k = k_1$,
	for which $\theta_0(k_1) = 1$,
	\begin{equation}
		k_1^2 = m^2 - \frac{g^2 \A^2}{2} - \frac{3 \epsilon}{8}
			\left( m^2 - \frac{g^2 \A^2}{8} \right) .
		\label{eq:kone-a}
	\end{equation}
	Collecting terms as before,
	it follows that the unstable matter modes grow at a rate
	determined by
	\begin{equation}
		\delta \chi_k \sim
		\exp\left[ 
			\left. \mu_k \right|_{max} \mstar t
			\left(1- \dfrac{3\epsilon}{16} \right)
		\right]
		\sim 
		\exp\left[
			\dfrac{1}{8} \left(\dfrac{\gstar\mathcal{A}}{\mstar}\right)^2
			\mstar t
			\left( 1+ \dfrac{3\epsilon}{16} \right)
		\right] .
	\end{equation}
	This displays a slight enhancement compared to the equivalent
	canonical model.

	\subsection{Preheating with a symmetry breaking term}
	\label{sec:sym-break}

	Now consider the symmetry-breaking potential
	given in Eq.~\eqref{eq:pot-sym}.
	It is convenient to redefine $\phi$ so that it
	describes excitations around the true vacuum at $\phi = \sigma$.
	After making the shift $\phi \rightarrow \phi+\sigma$, the
	potential can be written
	\begin{equation}
		V(\phi, \chi) = \dfrac{1}{2}m^2\phi^2
		+
		\dfrac{1}{2}g^2\phi^2\chi^2
		+
		g^2\sigma \phi \chi^2
		+
		\dfrac{1}{2}g^2\sigma^2\chi^2 .
		\label{eq:potentialbreaking}
	\end{equation}
	Excitations of $\chi$
	around this minimum acquire an effective mass
	$\meff^2\left(\phi\right)=m^2+g^2\chi^2$.

	Making the same assumption that preheating begins
	while $\chi \approx 0$,
	the inflaton modes will behave
	as in Eq.~\eqref{eq:resultinflaton}
	until the growth of $\chi$ occupation numbers
	causes the $\chi$ zero mode to grow.
	Until that time, the
	$\chi$ modes obey
	\begin{equation}
		\delta \ddot{\chi}_k
		+
		\left(
			k^2
			+ g^2\phi^2
			+ 2g^2\sigma\phi
			+ g^2\sigma^2
		\right)
		\delta \chi_k = 0 .
	\end{equation}
	We will suppose the brane executes oscillations around $\phi = \sigma$
	with amplitude $\A \ll \sigma$, so that it can reasonably be said
	to have settled at its final position.
	Assuming that $\sigma \ll m/g$,
	and again neglecting any bare mass for $\chi$,
	fluctuations in the matter field will be controlled by
	a Hill equation,
	\begin{equation}
		\delta \chi''_k
		+ \left[
			\theta_0
			+ 2 \theta_2 \cos 2z
			+ 2 \theta_6 \cos 6z
		\right] \delta \chi_k
		\simeq 0 ,
		\label{eq:hill-six}
	\end{equation}
	in which $2z=mt (1- 3 \epsilon / 16)$ and the parameters
	$\theta_0$, $\theta_2$ and $\theta_4$ obey
	\begin{equation}
		{
		\renewcommand{\arraystretch}{2}
		\begin{array}{lll}
			\theta_0 & = &
				\dfrac{4}{m^2}\left(k^2+ g^2\sigma^2\right)
				\left(1+\dfrac{3\epsilon }{8} \right) , \\
			\theta_2 & \simeq &
				4 \dfrac{g\sigma}{m} \dfrac{g\mathcal{A}}{m}
				\left(1+\dfrac{21\epsilon }{64} \right) , \\
			\theta_6 & \simeq &
				\dfrac{3\epsilon }{16} \dfrac{g\sigma}{m}
				\dfrac{g\mathcal{A}}{m} .
		\end{array}
		}
	\end{equation}
	To determine the growth of each $\chi$-mode we require an
	estimate of the Floquet exponent associated with Eq.~\eqref{eq:hill-six},
	which is again derived in Appendix~\ref{appendix:Hill}.
	For the $\ell = 1$ band, it is given by
	\begin{equation}
		\mu_k \simeq
			\sqrt{\left(\dfrac{\theta_2}{2}\right)^2\mp
				\dfrac{\theta_2\theta_6^2}{32} \mp
				\left(\dfrac{\theta_2}{2}\right)^3
			- \left(\theta_0^{1/2} -1 \right)^2} .
		\label{eq:floquet-six}
	\end{equation}
	As above, the terms involving $\theta_6^2$ and $\theta_2^3$
	can be neglected. The result is equivalent to a Mathieu equation
	with coefficients $A_k$ and $q$ satisfying
	\begin{equation}
		A_k = \dfrac{4}{m^2}\left(k^2+ g^2\sigma^2\right)
			\left(1+\dfrac{3\epsilon}{8} \right)
		\quad \mbox{and} \quad
		q = \theta_2 \simeq 4 \dfrac{g\sigma}{m} \dfrac{g\mathcal{A}}{m}
			\left(1+\dfrac{21\epsilon}{64} \right) \ll A_k .
		\label{eq:mathieu3}
	\end{equation}
	Since $q$ is much smaller than $A_k$, we expect resonance to be
	of the narrow type if it occurs at all.
	The first instability band
	occurs when $A_k = 1$
	and is centred at wavenumber $k = k_1$, where
	\begin{equation}
		k_1 \simeq \dfrac{m}{2} \left(1- \dfrac{3\epsilon}{16} \right) ,
		\label{eq:kone-b}
	\end{equation}
	which is shifted
	by $\Or(\epsilon)$ in comparison with the canonical case.
	Eqs.~\eqref{eq:kone-a} and~\eqref{eq:kone-b} show that the
	$\ell = 1$ resonance injects energy into a narrow band of
	modes with momenta $k \sim m$.
	Kofman, Linde \& Starobinksy
	interpreted the narrowness of this
	resonance as multiple, rapid copies of a
	decay process in which a single inflaton particle
	of mass $\sim m$ decays into two
	$\chi$-particles with opposite momenta
	of magnitude
	$\sim k_1$ \cite{Kofman:1996mv}.
	The width of the resonant band satisfies
	\begin{equation}
		\Delta k_1 \simeq
		2m \dfrac{g\sigma}{m} \dfrac{g\mathcal{A}}{m}
		\left( 1+ \dfrac{9\epsilon}{64}\right) .
	\end{equation}
	Using the appropriate limit of Eq.~\eqref{eq:floquet-six} to estimate
	$\mu_k$,
	or applying~\eqref{eq:mumathieu},
	we conclude that there is a small enhancement of preheating
	efficiency owing to the relativistic DBI correction.
	Near the centre of the $\ell = 1$ band, the $\chi$-modes
	grow like
	\begin{equation}
		\delta \chi_k \sim
			\exp\left(\left.\mu_k\right|_{\mathrm{max}} z \right)
		\sim
			\exp\left[\dfrac{\gstar\sigma}{\mstar}
			\dfrac{\gstar\mathcal{A}}{\mstar} \mstar t
			\left(1+ \dfrac{9\epsilon}{64} \right)\right] ,
	\end{equation}
	in which the growth rate is augmented with respect to
	the non-relativistic or canonical cases by
	an $\Or(\epsilon)$ term.

	\subsection{Efficiency of narrow-resonance preheating}

	The resonances we have discussed above are narrow, and
	therefore occupation numbers grow only for a limited set of
	momenta. This process is much less efficient than broad resonance,
	where a large range of wavenumbers experience growing occupation
	numbers. Nevertheless, because growth is exponential within the
	unstable bands,
	preheating by narrow resonance may
	still lead to an acceptably rapid conversion of inflaton modes
	into matter species. Gravitational redshift
	will cause
	modes of fixed comoving wavenumber
	to drift slowly through each unstable band from ultraviolet
	to infrared, and
	significant particle production will occur only when modes
	remain within an unstable band for sufficiently long
	that multiple e-foldings of stimulated emission occur \cite{Kofman:1997yn}.
	If wavenumbers move through the unstable bands too rapidly
	then little stimulated emission occurs, and the resonance
	is washed out.

	The arguments of Floquet's theory show that in each resonant band,
	the inflaton and matter fluctuations of wavenumber $k$
	grow like $\e{\mu_k z}$.
	Let us focus on the broadest instability band,
	which corresponds to $\ell = 1$
	and will generally dominate the particle production.
	This band has width $\sim q$, so each wavenumber remains within
	the unstable band for a time of order $q/H$.
	The resonance will be stable against Hubble expansion if
	$\left(\mu_k z/t\right) q / H \gtrsim 1$.
	It follows that preheating will drain energy from the
	inflaton provided
	\begin{equation}
		\sqrt{\dfrac{2H}{\mstar}} \lesssim \dfrac{3\epsilon}{8} .
	\end{equation}
	For the matter field, preheating will operate
	in the cases $\sigma = 0$ and $\sigma \neq 0$ whenever
	\begin{equation}
		\sqrt{\dfrac{2H}{\mstar}} \lesssim
			\dfrac{1}{4} \left( \dfrac{\gstar \mathcal{A}}{\mstar} \right)^2
			\left( 1 + \dfrac{13\epsilon}{32} \right)
	\end{equation}
	or
	\begin{equation}
		\sqrt{\dfrac{H}{\mstar}} \lesssim 
			2\left( \dfrac{\gstar \sigma}{\mstar}\right)
			\left( \dfrac{\gstar \mathcal{A}}{\mstar}\right)
			\left( 1+\dfrac{23\epsilon}{64}\right) ,
	\end{equation}
	apply, respectively.

	\section{Perturbative reheating}
	\label{sec:pr}

	While preheating is occurring it remains possible for
	inflaton particles to decay into other species via perturbative
	processes \cite{Albrecht:1982mp}.
	Indeed, these processes are necessary to thermalize the
	post-inflationary universe and bring reheating to an end.

	In the theories we have been considering,
	such decays are allowed by the symmetry-breaking potential
	studied in \S\ref{sec:sym-break},
	where there is an effective interaction $g^2 \sigma \phi \chi^2$
	in the true vacuum which permits the decay $\phi \rightarrow \chi \chi$.
	The simple potential~\eqref{masterequation}
	allows only two-body scattering and annihilation
	processes in vacuum.
	However, in the presence of an oscillating
	background field we can
	take the decay rate to be similar to that produced by
	the $g^2 \sigma \phi \chi$ interaction with $\sigma \sim \A$.
	If the inflaton has direct couplings to fermionic species then
	perturbative production can proceed via Yukawa interactions
	such as $h \phi \bar{\psi} \psi$, where $\psi$ is a Dirac fermion
	and $\bar{\psi} = \im \psi^\dag \gamma^0$ its adjoint.
	In this case we must suppose that any interactions in which the
	inflaton participates are not sufficient to spoil the
	possibility of successful inflation.
	
	In this section we will continue to assume that the inflaton mass
	is much larger than the bare mass of $\chi$, or the mass of any
	fermionic species where such couplings exist.
	Under these conditions, the tree-level decay rates
	for $\phi \rightarrow \chi \chi$ and $\phi \rightarrow \bar{\psi} \psi$
	can be calculated straightforwardly.
	The necessary expressions were given by Kofman {\etal}%
		\footnote{Modifications to these rates and cross-sections
		in a DBI model are studied in
		Appendix~\ref{appendix:canonicnorm}.}
	\cite{Kofman:1997yn}
	\begin{equation}
		\Gamma(\phi\to\chi\chi)
			= \frac{\gstar^4\sigma^2}{8\pi \mstar}
		, \quad \mbox{and} \quad
		\Gamma(\phi\to\bar\psi\psi)
			= \frac{\hstar^2 m}{8\pi} ,
		\label{eq:candecay}
	\end{equation}
	where $h_\star$ is the measured Yukawa coupling, determined by
	matching Eq.~\eqref{eq:candecay} to observation.
	Can perturbative decays give a contribution which is competitive
	with preheating?
	In one oscillation, perturbative decays populate the $\chi$ field
	at a rate $\Gamma/\mstar$. In the same time interval preheating
	increases the population at a rate $\mu_k$.
	Therefore perturbative decays dominate if
	\begin{equation}
		\Gamma \gtrsim \mstar \mu_k .
		\label{eq:reheating}
	\end{equation}
	Let us suppose that perturbative decays dominate the production
	of matter species.
	These processes are relatively slow
	and it is reasonable to assume that the decay products thermalize.
	After thermalization has gone to completion we can expect the
	resulting radiation dominated universe to have a temperature $T_R$,
	defined by
	\begin{equation}
		T_R \simeq 0.2 \sqrt{\Gamma M_P} ,
	\end{equation}
	where $\Gamma$ is the rate of the dominant decay channel. 

	Let us briefly recapitulate the role of perturbative processes in
	canonical inflation.
	Inflaton decays $\phi \rightarrow \chi \chi$
	can become the dominant channel for inflaton decay
	only in the case of symmetry breaking, for which $\sigma \neq 0$
	\cite{Kofman:1997yn}.
	These decays begin to dominate when the amplitude of inflaton
	oscillations has dropped to
	$\mathcal{A} \sim \gstar^2 \sigma / 8\pi$.
	If we suppose that the curvature perturbation synthesized during
	the inflationary phase, denoted $\zeta$,
	was dominated by fluctuations in
	the inflaton, then we must conclude $m \sim 10^{-5} \Mp$
	in order to match observation. This assumption is minimal but
	not mandatory; if $\zeta$ receives a dominant contribution
	from other sources, a wider range of masses $m$ can be tolerated.
	Supposing further that the coupling $\gstar$ has a natural value
	$\gstar \sim 10^{-1}$, we find a reheating temperature of order
	the symmetry breaking scale
	\begin{equation}
		T_R \simeq \sigma .
	\end{equation}

	Alternatively, it may happen that
	the dominant decay is into fermions.
	If so, Eq.~\eqref{eq:reheating} implies that
	the Yukawa coupling must satisfy
	\begin{equation}
		\hstar^2 \gtrsim 18\pi \left( \dfrac{\gstar\mathcal{A}}{\mstar} \right)
			\left( \dfrac{\gstar \sigma}{\mstar} \right) .
	\end{equation}
	Suitable values for the inflaton mass
	$\mstar$ and couplings $\{ \gstar, \hstar \}$
	can be found so that this condition applies, but these choices
	are model-dependent.
	As an example, if we assume
	$\hstar \sim 10^{-1}$ and $\mstar \sim 10^{-5} M_P$, we find the universe
	reheats to a temperature
	\begin{equation}
		T_R \simeq 1.3 \times 10^{-5} M_P
		\approx 3.12 \times 10^{13} \; \mathrm{GeV} .
	\end{equation}

	\subsection{Reheating temperatures}
	The above analysis can be applied at once to the
	production mechanisms discussed in
	\S\S\ref{sec:inf-prod}-\ref{sec:sym-break}.
	Let us suppose that $\sigma \neq 0$, so that we are working
	with the symmetry-breaking potential.
	Assuming a typical coupling $\gstar \sim 1$,
	Eq.~\eqref{eq:reheating} implies that
	$\phi \rightarrow \chi\chi$ decay dominates resonant production
	when the amplitude becomes of order
	\begin{equation}
		\mathcal{A} \lesssim 4\times 10^{-4} \sigma
		\left(1-\dfrac	{25\epsilon}{64}  \right)
		\label{eq:decay1}
	\end{equation}
	or smaller.
	The resulting reheating temperature receives relativistic
	corrections in comparison with
	the temperature achieved using canonical kinetic terms:
	\begin{equation}
		T_R \simeq 0.13 \sigma \, \left(1-\dfrac{\epsilon}{8}  \right) .
	\end{equation} 
	On the other hand, decays into fermions will become predominant
	($\phi \rightarrow \psi \bar{\psi}$)
	if the associated Yukawa coupling is sufficiently large
	\begin{equation}
		\hstar^2 \gtrsim 8\pi\, \left(\dfrac{\gstar \sigma}{\mstar} \right)
		\left(\dfrac{\gstar\mathcal{A}}{\mstar} \right)
		\left( 1+\dfrac{57\epsilon}{64} \right) .
		\label{eq:decay2}
	\end{equation}
	This is compatible with the conditions
	$g\mathcal{A}/m \ll g\sigma/m \ll 1$. In this case, we find that, for a
	representative value of the coupling, say $\hstar \sim 0.1$
	\begin{equation}
		T_R \simeq  3.12 \times 10^{13}
			\left(1-\dfrac{3\epsilon}{8} \right) \; \mathrm{GeV} .
	\end{equation} 

	\section{Conclusions}
	\label{sec:conclude}

	We have studied mechanisms by which the universe can be repopulated
	with matter species following an era of D-brane inflation.
	We work in the limit where the motion of the brane is at most
	perturbatively relativistic, which may ultimately be required by
	observation if $\fnl$ measured on equilateral configurations of
	the bispectrum is nonnegative.

	Our analysis is valid in a regime where the amplitude of oscillations
	is significantly smaller than
	whatever infrared scale caps the throat in
	which the D-brane moves.
	Indeed,
	because we take the D-brane to be at most perturbatively relativistic,
	inflation is of the slow-roll type.
	We assume inflation ends
	as the D-brane gently coasts to its resting position,
	dissipating its remaining energy by executing small amplitude
	oscillations. The amplitude and phase of these oscillations are
	mildly perturbated in comparison with the oscillations of
	canonically normalized scalar reheating the universe after a phase
	of four-dimensional inflation.

	We encounter some novel effects which are not present in a theory
	with canonical kinetic terms.	
	In canonical inflation there is no resonant production of
	inflaton particles during the reheating era.
	By contrast, in the perturbatively-relativistic
	limit of DBI inflation, non-linearities arising from the
	Dirac--Born--Infeld action cause a weak resonance
	which reaches an $\Or(1)$ effect only after
	$\sim 1/\epsilon$ oscillations.
	Excessive production of inflaton particles is typically
	problematic,
	but since preheating typically concludes after $\lesssim 10$
	oscillations this weak resonance is unlikely to be fatal for the
	DBI model. In the ultra-relativistic limit where $\gamma \gg 1$,
	the potential becomes unimportant and presumably
	resonance will be strongly suppressed.
	
	Another novel effect is a `DBI friction,' which appears in the
	equation of motion for the inflaton. This slightly damps the preheating
	effect. Beyond perturbation theory, it is possible this effect
	represents the decreasing significance of the potential at large $\gamma$.
	
	For a given set of measured masses and couplings which characterize
	the properties of the inflaton and the matter field into which it decays,
	we find that
	final reheating temperatures (assuming they are achieved
	by perturbative decays) are typically smaller
	than in the equivalent theory with canonical kinetic terms. 
	This may be beneficial in concrete models,
	where lower reheating temperatures allow problems associated
	with overclosure of the universe by moduli or gravitinos to be ameliorated.
	The most stringent constraint on DBI models comes from observations
	of the microwave background bispectrum, which presently require
	$\gamma \lesssim 18$. If our results are representative,
	and the reheating temperature falls as $\gamma$ increases,
	a constraint on $\gamma$ may also emerge from demanding
	that the nucleosynthesis era is uninterrupted, which requires
	$T_{\mathrm{RH}} \gtrsim \mbox{MeV}$.

	\acknowledgments

	NB is supported by the Wiener-Anspach Foundation.
	NB and DS acknowledge hospitality from DAMTP.
	ACD and DS acknowledge support from STFC.
	RHR is supported by FCT (Funda\c{c}\~{a}o para a Ci\^{e}ncia e a
	Tecnologia -- Portugal) through the grant SFRH/BD/35984/2007.
	We would like to thank Xingang Chen and David Tong for helpful
	discussions.

	\appendix

\section{Canonically normalized theory}
\label{appendix:canonicnorm}

	Consider a general higher-derivative Lagrangian
	of the form $P(\phi, X)$, given by Eq.~\eqref{eq:action}.
	We allow small excitations $\delta\phi$ around the background
	field $\phi_0$. Assuming derivatives of $\phi_0$
	are negligible there is a corresponding perturbation
	in the kinetic energy,
	$X = \delta\dot{\phi}^2 - (\partial{\delta \phi})^2$.
	To quadratic order in $\delta \phi$ but including all orders
	in other fields, the field propagates in a locally
	Minkowski region according to the action
	\begin{equation}
		S =
			\int \flatd^4 x \;
			\left\{
				P_0 +\dfrac{\partial P}{\partial \phi} \delta \phi
				+ \dfrac{\partial P}{\partial X} \left[
					\delta \dot{\phi}^2 - (\partial{\delta \phi})^2
				\right]
				+ \dfrac{1}{2} \dfrac{\partial^2 P}{\partial\phi^2}
					\delta\phi^2
				+ \cdots
			\right\} ,
		\label{eq:action:decay1}
	\end{equation}
	where $P_0 = P(X=0, \phi_0)$
	and `$\cdots$' denotes terms of higher order in $\delta \phi$
	and its derivatives which have been omitted.
	Eq.~\eqref{eq:action:decay1}
	will give a reliable description of scattering and
	decay processes involving inflaton particles
	provided derivatives of the background
	field are negligible over the relevant time and distance scales.
	In this limit we can neglect the curvature of spacetime
	and work in a locally flat region, 
	which should be chosen to be somewhat larger
	than the interaction region.
	Eq.~\eqref{eq:action:decay1} applies within this patch.
	S-matrix elements obtained from it
	can be expected to describe local
	scattering probabilities. On the other hand,
	if the background field $\phi$ varies significantly
	over the timescale of the scattering event then the definition
	of asymptotic \emph{in}- and \emph{out}-states becomes more complicated,
	and the concept of well-defined particles
	with an associated S-matrix may be invalidated.%
		\footnote{For example, this may happen in models where $\phi_0$
		evolves at a constant rate, such as the ghost inflation
		proposal of Arkani-Hamed {\etal}
		\cite{ArkaniHamed:2003uy,ArkaniHamed:2003uz}.
		However, this model is believed to suffer from
		certain pathologies.}
	
	\paragraph{Canonical normalization.}
	If the vacuum is locally stable on timescales comparable with
	decay processes, then $\langle \partial P/\partial \phi \rangle$ must be
	negligible when evaluated there. We must nevertheless
	retain perturbative
	excitations contained in this term, which include the fermion
	Yukawa couplings.
	After canonical normalization of $\delta \phi$,
	the interaction terms in the Lagrangian arise from
	\begin{equation}
		\mathcal{L}_{\mathrm{int}} \supseteq
			\frac{P_{,\phi}}{\sqrt{2 P_{,X}}} \delta \phi
			+ \frac{1}{4} \frac{P_{,\phi\phi}}{P_{,X}} \delta \phi^2 .
	\end{equation}
	where a comma denotes a partial derivative
	evaluated in the background.
	In a cut-off throat,
	the models we have been considering
	correspond to $P =- f^{-1} (1 - f X)^{1/2} + f^{-1} - V(\phi)$ with
	$f$ constant and $V(\phi)$ chosen to be
	Eq.~\eqref{masterequation} or~\eqref{eq:pot-sym}.
	Consider the symmetry-breaking potential~\eqref{eq:pot-sym}.
	We only wish to canonically normalize $\delta \phi$, which represents
	fluctuations around the mean field visible to local observers as
	particles.
	Therefore, one should shift to canonical normalization
	only after expanding in powers of fluctuations
	around the background field.
	Applying this procedure implies that we should make the transformations
	\begin{equation}
		m^2 \rightarrow \mstar^2 = \frac{m^2}{2 P_{,X}}
		\quad \mbox{and} \quad
		g^2 \rightarrow \gstar^2 = \frac{g^2}{\sqrt{2 P_{,X}}} .
		\label{eq:paramshifts}
	\end{equation}
	
	\paragraph{Measurable masses and couplings.}
	In this paper, we study the dynamics of particle production
	in different theories. If we are to make a meaningful comparison,
	these must be theories of the \emph{same} particles. Therefore
	our results should be written in terms of the masses and couplings
	measured, for example, by a local observer who can record the outcome of
	scattering events.
	
	In the case of non-canonical kinetic terms, it is possible that
	evolution of the background field causes the masses and couplings
	to change in time. In this case, it is the masses and couplings
	measured by a local observer at the time of reheating which are
	the relevant quantities.
	
	After canonically normalizing the fluctuation $\delta \phi$,
	and working with tree-level scattering amplitudes, there are no
	further rescalings in the LSZ formula. Therefore, at tree-level,
	the coupling $\gstar$ of the canonically normalized field is
	the coupling which would be measured by experiment.
	Likewise, the mass $\mstar$ of the canonically normalized field
	determines the position of the pole in its propagator.
	It is therefore the quantities $\mstar$ and $\gstar$ which should
	be compared with the mass and coupling in a theory with canonical
	kinetic terms.
		
	If a Yukawa coupling to fermions is present than
	we should similarly rescale its coupling constant,
	\begin{equation}
		h \rightarrow \hstar = \frac{h}{\sqrt{2 P_{,X}}} .
	\end{equation}

	\section{Floquet exponents from Hill's equation}
	\label{appendix:Hill}

	In this appendix we derive the Floquet exponent associated
	with Hill's equation.

We start with the case when $\theta_2, \theta_4 \neq 0$, and $\theta_{2n}=0$, for $n>2$ and by noting that in the edges of the resonance bands $\mu_k=0$; it turns out that for the most important instability band and for $\theta_2 \ , \ \theta_4 \ < \ 1$
\begin{displaymath}
\theta_0 \simeq 1 \pm \theta_2 \, \left[1\mp \dfrac{\left(\theta_4\right)^2}{6\, \theta_2} \right] \ \ ,
\end{displaymath} 
which, for $\theta_4 ^2 /\theta_2\ll 1$, essentially reduces to 
\begin{equation}
\theta_0 \simeq 1 \pm \theta_2 \ \ \ .
\label{eq:theta0}
\end{equation} 
From here, it also follows that, at lowest order, $\theta_0^{1/2} \simeq 1\pm \theta_2/2$. Furthermore, the instability parameter is given by \cite{McLachlan}
\begin{equation}
\sin^{2}\left({i\mu_{k} \frac{\pi}{2}}\right)=\triangle(0) \, \sin^{2}\left({\theta_0^{1/2}  \, \frac{\pi}{2}}\right) \ \ ,
\label{eq:mu}
\end{equation}
where $\triangle(0)$ is the discriminant evaluated for $\mu=0$ (following from Hill's method); $\triangle (0)$ is exclusively determined by the parameters $\theta_0$, $\theta_2$ and $\theta_4$  and is approximately given by
\begin{displaymath}
\triangle(0)\simeq 1+\dfrac{\pi}{4\, \theta_0^{1/2} }\, \cot{\left(\theta_0^{1/2} \, \dfrac{\pi}{2}  \right) \, \left[ \dfrac{\theta_2\, ^2}{1-\theta_0}  +  \dfrac{\theta_4\, ^2}{4-\theta_0}      \right]} \ \ ;
\end{displaymath} 
it can be shown that this, approximately, yields 
\begin{equation}
\triangle(0) \simeq 1\mp \left(\dfrac{\pi}{2}\right)^2\, \left[\mp \left(\dfrac{\theta_2}{2}\right)^2 +\dfrac{\theta_2}{3}\, \left(\dfrac{\theta_4}{2}  \right)^2 +\dfrac{\left(\theta_2  \right)^3}{2}     \right]  \ \ .
\label{eq:triangle0}
\end{equation} 
Using (\ref{eq:theta0}) and (\ref{eq:triangle0}), we find that from (\ref{eq:mu}), we may arrive at an approximate expression for the instability parameter for this Hill's equation as 
\begin{equation}
\mu_k \simeq \sqrt{\left(\dfrac{\theta_2}{2}\right)^2 -\left(\theta_0^{1/2}  -1 \right)^2  \mp \dfrac{\theta_2}{3}\, \left(\dfrac{\theta_4}{2}  \right)^2      \mp  \dfrac{\left(\theta_2  \right)^3}{2}     } \ \ .
\end{equation} 
This expression for $\mu_k$ is valid for $\theta_2, \theta_4 \ll 1$.

If we now proceed and consider the case with non-vanishing $\theta_0$, $\theta_2$ and $\theta_6$, we observe that the instability parameter may be computed using again using (\ref{eq:mu}), where now 
\begin{equation}
\Delta(0) \simeq 1+ \dfrac{\pi \cot\left(\dfrac{\pi}{2}\, \theta_0^{1/2}  \right)}{4\theta_0^{1/2}} \ \left( \dfrac{\theta_2^2}{1-\theta} +\dfrac{\theta_6^2}{9-\theta_0}  \right) \ \ .
\label{eq:defindelta}
\end{equation} 
By considering the first instability band, we take $\theta_0 \simeq 1 \pm \theta_2$, so that
\begin{equation}
\theta_0^{1/2}\simeq 1 \pm \dfrac{\theta_2}{2} \ \ , \ \ \dfrac{\theta_2^2}{1-\theta_0} = \mp \theta_2  \ \ ,\ \ \dfrac{\theta_6^2}{9-\theta_0} \simeq \mp \dfrac{\theta_6^2}{8}\, \left(1 \pm \dfrac{\theta_2}{8} \right) \ \    \textrm{and} \ \ \cot\left(\dfrac{\pi}{2}\, \theta_0^{1/2} \right) \simeq \mp \dfrac{\pi}{4}\, \theta_2  \ \ .
\end{equation} 
Plugging these into (\ref{eq:defindelta}), we find that, up to lowest order
\begin{displaymath}
\Delta (0) \simeq 1 \mp \left(\dfrac{\pi}{4}\right)^2 \ \left(\mp \theta_2^2 +\dfrac{\theta_2\theta_6^2}{8} +\dfrac{\theta_2^3}{2}  \right) \ \ .
\end{displaymath} 
Moreover, 
\[\sin\left(\dfrac{\pi}{2}\, \theta_0^{1/2} \right) \simeq 1-\dfrac{1}{2}\, \left( \dfrac{\pi}{2}\right)^2\, \left(\theta_0^{1/2}-1 \right)^2 \ \ .\]
Finally, by substituting into (\ref{eq:mu}), we obtain 
\begin{equation}
\mu_k \simeq \sqrt{\left(\dfrac{\theta_2}{2}\right)^2\mp \dfrac{\theta_2\theta_6^2}{32} \mp \left(\dfrac{\theta_2}{2}\right)^3 -\left(\theta_0^{1/2} -1 \right)^2     } \ \ .
\end{equation} 

\providecommand{\href}[2]{#2}\begingroup\raggedright\endgroup

\end{document}